\documentclass[runningheads]{llncs}

\usepackage[T1]{fontenc}
\usepackage{graphicx}
\usepackage[T1]{fontenc}
\usepackage{cite}
\usepackage{amsmath,amssymb,amsfonts}
\usepackage{graphicx}
\usepackage{textcomp}
\usepackage{xcolor}
\usepackage[many]{tcolorbox}
\usepackage{booktabs}
\usepackage{hyperref}

\usepackage[many]{tcolorbox}
\usepackage{xcolor}
\usepackage{tikz,pifont}
\usepackage{amsfonts,amsmath}
\usepackage{makecell}
\usepackage{xspace}
\usepackage{algorithm,algpseudocode}
\usepackage{listings}
\usepackage{makecell,colortbl}

\makeatletter
\let\MYcaption\@makecaption
\makeatother

\usepackage[font=footnotesize]{subcaption}

\makeatletter
\let\@makecaption\MYcaption
\makeatother

\newcommand*{\ie}{i.e.,\@\xspace}
\newcommand*{\eg}{e.g.,\@\xspace}

\newcommand*{\etal}{\emph{et~al.}\@\xspace}

\newcommand{\rqsecond}{\textbf{RQ$_1$}: \emph{To what extent an optimal hyperparameter setting is effective for different domain models?}\xspace}

\newcommand{\rqthree}{\textbf{RQ$_2$}: \emph{How the combination of hyperparameter tuning and prompt engineering techniques can improve the quality of models generated for different domains?}\xspace}

\newcommand{\lifemap}{LIFEMap\xspace}

\newtcolorbox{rqanswer}{
    colback=gray!25, 
    colframe=black, 
    enhanced,
    boxrule=0mm,
    frame hidden,
    left=1mm, 
    top=1mm, bottom=1mm, right=1mm, 
    borderline west={0.8mm}{0mm}{gray!80!black}, 
    sharp corners, 
    fontupper=\small 
}

    \lstdefinestyle{searchstringstyle}{
	basicstyle=\ttfamily\scriptsize,
	captionpos=b,                    
	numbers=none,                    
	numbersep=4pt,                  
	showspaces=false,                
	showstringspaces=false,
	showtabs=false,                  
	tabsize=2,
	frame=single
}

\begin{document}

\title{Investigating the Role of LLMs Hyperparameter Tuning and Prompt Engineering to Support Domain Modeling}
\titlerunning{Investigating LLMs Tuning and Prompt Engineering for Domain Modeling}
%
\author{
Vladyslav Bulhakov \and Giordano d'Aloisio\orcidID{0000-0001-7388-890X} \and Claudio {Di} Sipio\orcidID{0000-0001-9872-9542} \and Antinisca {Di} Marco\orcidID{0000-0001-7214-9945} \and Davide {Di} Ruscio\orcidID{0000-0002-5077-6793}   
}
\authorrunning{V. Bulhakov et al.}
%
\institute{DISIM Department, University of L'Aquila, L'Aquila, Italy\\
\email{vladyslav.bulhakov@student.univaq.it \{giordano.daloisio,claudio.disipio,antinisca.dimarco,davide.diruscio\}@univaq.it}}
\maketitle              
\begin{abstract}
The introduction of large language models (LLMs) has enhanced automation in software engineering tasks, including in Model Driven Engineering (MDE). However, using general-purpose LLMs for domain modeling has its limitations. One approach is to adopt fine-tuned models, but this requires significant computational resources and can lead to issues like catastrophic forgetting.

This paper explores how hyperparameter tuning and prompt engineering can improve the accuracy of the Llama 3.1 model for generating domain models from textual descriptions. We use search-based methods to tune hyperparameters for a specific medical data model, resulting in a notable quality improvement over the baseline LLM.
We then test the optimized hyperparameters across ten diverse application domains. 

While the solutions were not universally applicable, we demonstrate that combining hyperparameter tuning with prompt engineering can enhance results across nearly all examined domain models.

\keywords{domain modeling \and large language models \and MDE}
\end{abstract}

\section{Introduction}
The interplay of Model Driven Engineering (MDE) and large language models (LLMs) has recently gained popularity, leading to different applications in several modeling tasks, including model completion \cite{chaaben2023towards}, model generation \cite{chen_automated_2023,arulmohan_extracting_2023}, and specification of software architectures \cite{ahmad_towards_2023}. Nevertheless, full automation in modeling tasks is far from being reached \cite{10.1145/3712008}, especially in domain modeling tasks, where human expertise is pivotal \cite{10.5555/993859}. Moreover, general-purpose LLMs exhibit limitations in assisting modelers in completing or generating domain models \cite{chen_automated_2023}. One practical solution is represented by fine-tuning processes when an LLM can be specialized for a particular task \cite{ma2024fine}. While this method has been proven to be effective in domain modeling \cite{10.1145/3640310.3674089}, it requires high-quality datasets that need to be collected, pre-processed, and encoded to train the model. Moreover, fine-tuned models can suffer from the so-called catastrophic forgetting \cite{pmlr-v139-chen21v}, \ie a significant drop in performance on previously learned tasks. Alternatively, the overall performance of a deep learning model can be increased by relying on the hyperparameter tuning technique \cite{nsgaII}, which is focused on finding the best combination of hyperparameters without additional training phases. 

In this paper, we investigate the effect of hyperparameter tuning and prompt engineering on a notable LLM \ie Llama 3.1 \cite{LLama3} when employed in a very specific task, i,e., domain modeling. First, we exploit the NSGA-II algorithm to find an initial set of hyperparameters for a specific application domain, \ie medicine, using one single domain model, \ie related to the \lifemap project.\footnote{\url{https://www.thelifemap.it/}} Afterward, a grid search is performed to refine the set of hyperparameters aiming at reaching a trade-off between semantic and syntactical correctness. In addition to the hyperparameter tuning, we experiment with three notable prompt techniques, \ie zero-shots, few-shots, and chain-of-thoughts, as they have been successfully used in domain modeling \cite{chen_automated_2023}. 

To evaluate our approach, we compare models generated by Llama 3.1 with optimal configurations to those produced by Llama 3.1 using default hyperparameters, both utilizing the \lifemap domain model as input. The performed experiments confirm that the models generated by the LLM with optimal configurations are significantly better than the baseline, motivating the need for hyperparameter tuning. Then, we assess the generalizability of the optimal solutions. To this end, we run the Llama 3.1 model using the identified list of hyperparameters on ten different application domains and collect the results for each prompt technique. Again, we compare the obtained results with a baseline model with default hyperparameters. While our results show that hyperparameter tuning alone is insufficient to achieve adequate automation in different domains, the combination with prompt technique helps mitigate this issue, achieving better results in most of the analyzed domains. These results again motivate the need for LLM's hyperparameter tuning to support domain model generation. In particular, hyperparameter tuning does not require a large set of modeling artifacts to be conducted, offering a valuable alternative when fine-tuning is not applicable. In future work, we plan to extend this analysis to other notable LLMs - like GPT, BERT, or DeepSeek - to assess if hyperparameter tuning is also effective with different models.

The main contributions of the paper are: \textit{i)} a rigorous
study on hyper-parameter tuning on \lifemap domain; \textit{ii)} an extensive evaluation with 10 state-of-the-art domain models to assess to what extent our approach is generic; \textit{iii)} a replication package to foster research in this domain \cite{repl_package}.

\label{sec:introduction}

\section{Motivation and Related Work}\label{sec:background}

\begin{figure}[tb]
    \centering
    \includegraphics[width=1\linewidth]{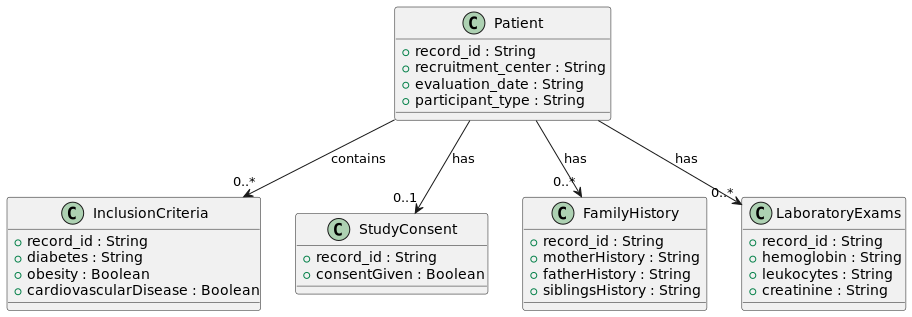}
    \caption{Excerpt of the \lifemap domain model}
    \label{fig:motivating}
\end{figure}

\noindent
\textit{Motivating example:} \label{sec:motivation}
Through domain modeling, formal representations typically specified in class diagrams are conceived out of textual requirements. Although several approaches have been developed to assist modelers in domain modeling, the process is still challenging due to the complexity of peculiar domains. Figure \ref{fig:motivating} depicts an excerpt of the \lifemap model, which represents a medical Case Report Form (CRF) for patients under specific treatments. In particular, a \texttt{Patient} has a \texttt{FamilyHistory} that may affect the treatment and the corresponding \texttt{LaboratoryExams}.
It is worth mentioning that the discussed model is only an excerpt of the whole domain model,\footnote{The complete model is available in the online appendix \cite{repl_package}} which contains  31 classes, 291 attributes, and 48 relationships. 

To produce domain models, modelers must manually inspect huge textual documents to extract relevant concepts. Relying on automated techniques, like intelligent modeling assistants \cite{mussbacher_opportunities_2020}, requires curated and large modeling datasets \cite{modelset}. Moreover, documents might contain sensitive data that cannot be disclosed, as modeled by the entity \texttt{StudyConsent}, thus limiting training data to perform advanced training techniques, \eg fine-tuning or retrieval augmented generation (RAG) \cite{lewis2020retrieval,10651492}. To overcome such limitations, in this paper, we leverage the hyperparameter tuning strategy by providing a set of hyperparameters that can support the usage of LLMs in this peculiar domain.

\medskip
\noindent
\textit{Related work:} DoMoBOT \cite{10.1145/3417990.3421385} combines NLP and a supervised ML model to automatically retrieve domain models from textual contents using the spaCy tool.\footnote{\url{https://spacy.io/}} After the pre-processing phase, the predictive component uses the encoded sentences to retrieve similar model entities and generate the final domain model.

In \cite{10.1145/2976767.2976785}, authors exploit the NSGA-II algorithm to retrieve partial metamodels. The system first searches a list of candidates given inputs relevant to the considered task. Then, several similarity criteria are selected as additional optimization objectives to support model completion. 



Chaaben \etal \cite{chaaben2023towards} supports model completion using GPT-3 equipped with using semantic mapping, \ie embedding the model elements as structured text in the prompt. Preliminary results computing traditional accuracy metrics on 30 models extracted from the ModelSet dataset~\cite{modelset} show that the few-shot approach can help modelers complete static UML models. 
In \cite{10.1145/3567512.3567534}, the authors exploited the GPT-3 model with a few-shots prompt technique to translate textual requirements in domain-specific modeling language (DSL) specifications in the context of Advanced Driver Assistance Systems (ADAS). Starting from unstructured textual requirements, they derive formal rules used in the DSL specification.

Camara \etal \cite{camara_assessment_2023} explored the capability of GPT3.5 to automate domain modeling tasks, including OCL rule generation. The conducted experiment, considering 40 models belonging to 8 different domains, shed light on limitations in model generation. 

Mutagene \cite{10.1145/3652620.3687808} is a custom GPT-4 model to support the development of DSMLs that relies on a dedicated knowledge base,
including the Xtext syntax, examples, and documentation to support a specific use case, \ie modeling mutation. 
Similarly, ModelMate \cite{10.1145/3640310.3674089} relies on fine-tuned LLM to assist modelers in three different DSL-related tasks, \ie identifier suggestion, line completion, and block completion. In particular, three different LLMs have been fine-tuned, \ie GPT-2, Code Parrot, and Code-Gen, to support three tasks, showing superior accuracy compared to traditional techniques considering the three examined tasks. 
A conceptual framework to integrate LLMs with graphical DSML has been proposed in \cite{10.1145/3640310.3674085}. First, the high-level concepts are extracted from the current literature. Then, a proof-of-concept using ChatGPT is discussed, \ie the proposed framework is used to design a simple traffic light DSML for individuals with color blindness. 

The most related work to ours is \cite{chen_automated_2023}, where the authors experimented with GPT-3.5 and GPT-4 using different prompt engineering methods and conducted a detailed comparative evaluation with precision, recall, and F1-measure, plus a manual comparison using a tailored correctness metric. While we rely on the same dataset to evaluate our approach, we cannot directly compare the obtained results as \textit{i)} we rely on a smaller LLM, \ie Llama 3, and \textit{ii)} we adopt Cosine and BERT scores as similarity functions in the hyperparameter tuning as they provide proper integration with the NSGA-II algorithm.


\section{Methodology}\label{sec:methodology}
This section outlines the procedures used to identify the optimal hyperparameter configurations for the LLM, as well as the prompt engineering strategies that were implemented on the considered LLM, \ie Llama 3.1.


\subsection{Hyperparameter Tuning}

\begin{figure}[tb]
    \centering
    \includegraphics[width=0.7\linewidth]{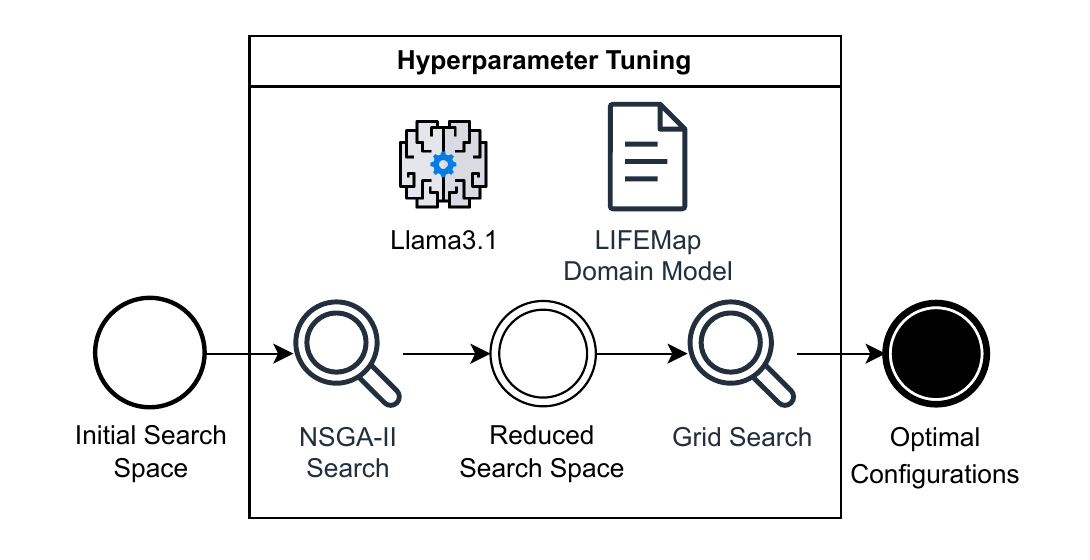}
    \caption{Hyperparameter tuning methodology}
    \label{fig:hyperparam}
\end{figure}

Figure \ref{fig:hyperparam} reports an overview of the hyperparameter tuning process. We start from an initial search space of hyperparameters and possible value combinations. Next, we employ a global search strategy - namely NSGA-II \cite{nsgaII} - to reduce the space of possible solutions. From this reduced search space, we perform an exhaustive grid search process to evaluate each potential solution. From this search process, we derive a set of optimal solutions that achieve the best semantical and syntactical correctness trade-off of the generated models. As mentioned above, the whole search process is performed using Llama3.1 as the base LLM. In particular, we ask the LLM to generate an Ecore metamodel of the \lifemap textual domain model outlined in Section \ref{sec:background}. 


\subsubsection{Search Space}\label{sec:search_space}

In our study, we fine-tune the following hyperparameters of Llama3.1, which have been shown to be the most influential on the quality of the generated content \cite{arora2024optimizinglargelanguagemodel}:

\medskip
\noindent \ding{228} \textbf{Temperature:} Influences output randomness; lower values (close to 0) make it deterministic, while higher values ($\geq$ 1) increase randomness. 

\noindent \ding{228}  \textbf{Top-k:} Determines the number of high-probability tokens to consider for text generation. Lower values (1-10) result in more deterministic behavior; higher values increase randomness.

\noindent \ding{228}  \textbf{Top-p:} Used with top-k to select tokens based on cumulative probability. Lower values (0.1-0.3) create a smaller set for selection, making the output more deterministic, while higher values (0.8-0.9) expand the selection set. 

\noindent \ding{228}  \textbf{Repetition penalty:} Limits excessive repetition in output. Values close to 1 allow repetition; values ($\geq$ 1.8) discourage it. 

\noindent \ding{228}  \textbf{Max new tokens:} Sets the maximum number of tokens that can be generated in response to a prompt, with higher values leading to longer responses.

\begin{table}[tb!]
    \centering
    \caption{Hyperparameters search space}
    \label{tab:values_range}
    \begin{tabular}{l|r}
    \toprule
        \textbf{Hyperparameter} & \textbf{Value Range} \\
    \midrule
    \midrule
       Temperature & $[0.5, 2] \;\; \text{with step size of } 0.1$ \\
       Top-k & $[0, 100] \;\; \text{with step size of } 10$\\
       Top-p & $[0.5, 1.0] \;\; \text{with step size of } 0.1$\\
       Repetition penalty & $[1.0, 2.0] \;\; \text{with step size of } 0.1$\\
       Max new tokens & $\{512, 1024, 2048, 3072, 4096, 8192\}$\\\bottomrule
    \end{tabular}
\end{table}

\medskip
Since there is no agreement on which range of hyperparameter values is more suited for model generation tasks \cite{chen_automated_2023,arora2024optimizinglargelanguagemodel}, we explore a wide range of values that spans from low to high settings. The search space is reported in Table \ref{tab:values_range}. 

Given the size of the search space and the high computational complexity of generative models, performing an exhaustive search of this space is infeasible. Thus, we rely on an advanced multi-objective search strategy to refine this space, as discussed in the next section.


\smallskip
\subsubsection{Search Strategies}\label{sec:search}

We employ a combination of two search strategies to explore the search space described in Section \ref{sec:search_space}. The objective of our search task is to identify hyperparameter configurations within the search space that allow the LLM to produce models with the highest levels of semantic and syntactic correctness compared to a ground truth model. Thus, this task can be defined as a multi-objective search problem where we maximize both the semantic and syntactic correctness of the generated models.

First, we rely on a global search strategy to refine the search space. Secondly, starting from the reduced search space, we perform a comprehensive grid search of all the possible combinations. Eventually, the search process returns the solutions that achieve the best trade-off of semantic and syntactical correctness for the generated models.

\smallskip
\paragraph{Global Search}

The global search strategy employed in our work is the NSGA-II algorithm \cite{nsgaII}. It is an evolutionary algorithm that has been shown to be widely efficient and effective for multi-objective problems \cite{gong2024greenstableyolo,sarro_multi-objective_2016,hamdani2007multi}. It follows the general behavior of genetic evolutionary algorithms, where an initial population of individuals progressively evolves following Darwin's theory of biological evolution \cite{forrest1996genetic}.


The algorithm starts with an initial population of $N$ randomly generated individuals, and the quality of each individual is assessed following a given \textit{fitness} function. Next, the evolution process begins. First, the best individuals from the population are selected following a given \textit{selection} strategy, creating an offspring population $O_t$. This new population is evolved using specific \textit{crossover} and \textit{mutation} strategies under specific probabilities $p_c$ and $p_m$, respectively, and the quality of the new individuals is again evaluated. The new offspring population $O_t$ is merged with the population of the $t$ generation $P_t$ to create a temporary population $P'$. The individuals from $P'$ are sorted based on a non-dominated sorting strategy. Finally, the first $N$ individuals are selected from $P'$ to create the new population $P_{t+1}$. This process is repeated until the maximum number of generations $G$ is reached and the final population $P^*$ is returned. 


In our use case, each individual in the search is represented as a dictionary of hyperparameters and related values, and the population size is 30. The number of generations is set to 10. The \textit{selection} strategy is a \textit{Tournament Selection} based on dominance between two individuals. The number of individuals selected is equal to the population size. The \textit{crossover} function is a \textit{Single Point Crossover} operator \cite{kora2017crossover} with 90\% of probability. The \textit{mutation} strategy is a Uniform Mutation operator where all hyperparameters are randomly mutated. The mutation probability is 20\%. Finally, the quality of each individual is evaluated using two fitness functions for syntactical and semantic similarity against a ground truth model. Concerning syntactic similarity, we employ the widely adopted \textbf{Cosine Similarity} score \cite{salton1988term}. 
This metric measures the similarity of two documents as the cosine of their angle in their term-frequency vector representation. If two documents are syntactically equal, then their cosine similarity is one. 
Instead, we use the \textbf{BERTScore} metric for semantic similarity. This metric evaluates the quality of a generated text by computing the similarity between the BERT embeddings of two documents \cite{zhang2019bertscore}. 
As for the Cosine Similarity, the optimal value of this metric is one. 

The choice for those metrics is driven by their employment in previous works related to domain modeling \cite{weyssow2022recommending,di2022memorec}. While we acknowledge that there are more specific metrics to assess the similarity of modeling artifacts, \eg EMFCompare \footnote{\url{https://github.com/eclipse-emf-compare/emf-compare}}, they do not provide an API to be used as a fitness function of the NSGA-II algorithm.

\begin{table}[tb!]
    \centering
    \scriptsize
    \caption{Reduced search space}
    \label{tab:small_space}
    \begin{tabular}{l|r}
    \toprule
        \textbf{Hyperparameter} & \textbf{Value Range} \\
    \midrule
    \midrule
       Temperature & $[1.0, 1.3] \;\; \text{with step size of } 0.1$ \\
       Top-k & $\{0, 50\}$\\
       Top-p & $[0.9, 1.0] \;\; \text{with step size of } 0.1$\\
       Repetition penalty & $\{1.0, 1.1, 1.2\}$\\
       Max new tokens & $\{512, 1024, 2048, 3072, 4096\}$\\\bottomrule
    \end{tabular}
\end{table}
As stated above, we employ this global search strategy to reduce the search space of possible solutions. In particular, we ran the abovementioned algorithm ten times and identified the range of values for each hyperparameter in the final populations. Notably, we obtained a smaller set of hyperparameters, which made it suitable to perform an exhaustive grid search. The reduced space of solutions is reported in Table \ref{tab:small_space}.   

\smallskip
\paragraph{Grid Search}

Starting from the reduced space of solutions returned by NSGA-II, we performed an exhaustive grid search by evaluating each possible hyperparameter setting in the search space. We conducted this additional search to evaluate all possible configurations and identify the optimal settings that could be reused for all domain modeling generation tasks. 

The final outcome is the Pareto front of the solutions with respect to BERT Score and Cosine Similarity. The Pareto front solutions are those that are not dominated by any other solution in the search space - \ie they are better than all solutions in at least one fitness score and no worse in all the others \cite{10.1162/EVCO_a_00128}. In particular, we identified six different configurations, which are the subject of our evaluation. The whole search process (\ie NSGA-II search plus Grid search) took around ten days of machine execution over a CentOS HPC cluster equipped with 32 Intel(R) Xeon(R) Gold 6140M CPUs and two Nvidia A100 and A30 GPUs. 
The optimal configurations are reported in Table \ref{tab:solutions}. 

\begin{table}[tb]
\centering
\scriptsize
    \caption{Optimal solutions and Llama default configuration}
    \label{tab:solutions}
   \begin{tabular}{l|c}
    \toprule
         & \textbf{Configuration} \\
    \midrule \midrule
       \textbf{S0} & \{Temperature: 0.6, Top-k: 50, Top-p: 1.0, Max New Tokens: 4096\}\\
       \midrule
       \textbf{S1} & \{Temperature: 1.0, Top-k: 0, Top-p: 1.0, Max New Tokens: 2048\}\\\midrule
       \textbf{S2} & \{Temperature: 1.1, Top-k: 50, Top-p: 0.9, Max New Tokens: 3072\}\\\midrule
       \textbf{S3} & \{Temperature: 0.8, Top-k: 50, Top-p: 0.9, Max New Tokens: 4096\}\\\midrule
       \textbf{S4} & \{Temperature: 1.1, Top-k: 50, Top-p: 0.9, Max New Tokens: 3072\}\\\midrule
       \textbf{S5} & \{Temperature: 1.1, Top-k: 50, Top-p: 0.9, Max New Tokens: 4096\}\\
       \midrule
       \midrule
       \textbf{Llama Default} & \{Temperature: 0.7, Top-k: 50, Top-p: 0.9, Max New Tokens: 4096\}\\
       \bottomrule
    \end{tabular}
\end{table}

\subsection{Prompt Engineering}

In this work, we analyze the three most adopted prompt engineering strategies for software engineering tasks. The choice of those strategies is also motivated by previous work on the adoption of LLMs for domain modeling \cite{chen_automated_2023}. In the following, we summarize each adopted prompt strategy, while the complete prompts are reported in our replication package \cite{repl_package}.

\subsubsection{Zero-shot}\label{sec:zero-shot}

This strategy is our baseline and has been employed during the hyperparameter tuning phase. An excerpt of the prompt employed is reported in Listing \ref{lst:zero}. In general, with a zero-shot prompt, the LLM is fed only textual instructions on the task to be performed and the input file to process. In our case, we give the LLM a textual description of the LIFEMap domain model and ask it to extract all entities, attributes, and relations from it. We ask the LLM to behave as a domain modeling expert and generate the output as an Ecore model.   

\begin{lstlisting}[caption=Excerpt of Zero-shot prompt,label=lst:zero,style=searchstringstyle]
You are a domain modeling expert.
Identify all entities in the text.
Identify all attributes and data types.
Identify all relationships among entities.
Generate the output as an Ecore model.
<textual LIFEMap domain model>
\end{lstlisting}

\subsubsection{Few-shot}\label{sec:few-shot}

Few-shot prompting extends the zero-shot strategy by including examples of the output that the LLM has to generate. In our case, we append two examples of Ecore models to the end of the prompt. An excerpt of the prompt is given in Listing \ref{lst:few}.  

\begin{lstlisting}[caption=Excerpt of Few-shot prompt,label=lst:few,style=searchstringstyle]
<Zero-Shot prompt>
<example ecore model 1>
<example ecore model 2>
<textual LIFEMap domain model>
\end{lstlisting}

\subsubsection{Chain-of-Thought}\label{sec:cot}

Chain-of-Thought (CoT) is an extension of the few-shot strategy that includes reasoning steps in addition to the examples provided to the LLM \cite{wei2022chain}. The reasoning steps link the examples provided with the task that the LLM has to solve. In particular, in addition to the two examples of Ecore models, we feed the LLM with textual descriptions of those example domain models and the steps to follow to define Ecore models from those textual descriptions. An excerpt of this prompt is given in Listing \ref{lst:cot}.

\begin{lstlisting}[caption=Excerpt of Chain-of-Thought prompt,label=lst:cot,style=searchstringstyle]
<Zero-Shot prompt>
<textual description of example 1>
<steps to derive ecore model for example 1>
<example ecore model 1>
<textual description of example 2>
<steps to derive ecore model for example 2>
<example ecore model 2>
<textual LIFEMap domain model>
\end{lstlisting}

Concerning the examined model, we focus on the \texttt{Llama-3.1-8B-Instruct} model available in the HuggingFace repository.\footnote{\url{https://huggingface.co/meta-llama/Llama-3.1-8B-Instruct}} By the time of this paper, Llama 3.1 is among the latest releases of the family of Llama language models that exploit optimized transformer architecture and reinforcement learning with human feedback. Thus, the model is widely adopted for many tasks, including code-related tasks \cite{10.1145/3709358,deroy2024code}.

\section{Evaluation}\label{sec:evaluation}
\subsection{Reseach questions}

\begin{itemize}

    
 \item \rqsecond This research question evaluates whether optimal hyperparameter configurations can be generalized across different domain models without prompt engineering. We begin with a sanity check using the \lifemap model to verify that the improved hyperparameters enhance the semantic and syntactic quality of generated models. We then apply these hyperparameters across various domain models using an existing dataset \cite{chen_automated_2023}. If the models are semantically and syntactically correct, we conclude that hyperparameter tuning can be done once, allowing for generalization without additional prompt engineering.

 \item \rqthree To answer this question, we experiment with advanced prompt engineering strategies, \ie Few-Shot, and CoT. Specifically, suppose the LLM, configured with a particular hyperparameter setting and employing a specific prompt engineering strategy, effectively generates models that are both semantically and syntactically correct across various domains. In that case, we can conclude that hyperparameter tuning can be performed once, and the optimal settings can be applied for multiple domain models when using that prompt engineering strategy.
    
\end{itemize}




\subsection{Dataset}

To assess the generalizability of the optimal solutions, we evaluate the semantical and syntactical correctness of domain models generated for the use cases employed in \cite{chen_automated_2023}.
In particular, the dataset includes ten heterogeneous domain models:
\textbf{BTMS:} \textit{transportation} domain; \textbf{H2S:} \textit{food delivery} domain; \textbf{LabTracker:} \textit{health} domain; \textbf{CeIO:} \textit{social} domain; \textbf{TSS:} \textit{sport} domain; \textbf{SHAS:} \textit{IoT} domain; \textbf{OTS:} \textit{teaching} domain; \textbf{Block:} \textit{game} domain; \textbf{Tile-O:} \textit{game} domain; \textbf{HBMS:} \textit{managment} domain.     

A more detailed description of these domain models is available in \cite{chen_automated_2023}. However, it is worth noting that they have different degrees of complexity regarding the number of classes, attributes, and relations. The number of entities ranges from 7 to 23. The attributes span from 11 to 43, whereas the number of relationships ranges from 9 to 27. Nevertheless, none of them is comparable with the \lifemap model in terms of size as discussed in Section \ref{sec:motivation}.

\subsection{Prompt engineering and metrics}

For answering RQ$_1$, we use a zero-shot prompting strategy, maintaining the same structure as illustrated in Section \ref{sec:zero-shot} that represents our baseline model. To answer  RQ$_2$, we utilize prompts designed with few-shot and chain-of-thought strategies, adhering to the structures detailed in Sections \ref{sec:few-shot} and Section \ref{sec:cot}, respectively. Note how, for RQ$_2$, the prompt engineering strategies are also used by the baseline LLM with default hyperparameters.

As mentioned in Section \ref{sec:methodology}, the syntactic correctness is evaluated by computing the Cosine Similarity between the generated models and a ground truth consisting of a domain model manually created \cite{salton1988term}. The semantic similarity is instead assessed by computing the BERTScore \cite{zhang2019bertscore} between the generated models and the ground truth. These metrics are employed in both RQs.

Finally, to account for the randomness of LLMs in output generation, following previous works \cite{chen_automated_2023,d2024exploring}, for each RQ, we repeat the model generation process 20 times for each prompt and hyperparameter setting. In addition, we perform the one-sided non-parametric Wilcoxon test to assess if there is a statistically significant difference in the results obtained for each RQ \cite{woolson2005wilcoxon}. 
Following common methodology \cite{wohlin_experimentation_2012}, if the test's $p$-value is $\leq 0.05$, then we can reject the null hypothesis and assess, with a statistical significance, that the Cosine Similarity and BERT Score of the domain models generated by the LLM with optimal configurations are higher (\ie better) than the baseline. We chose this test instead of the paired \textit{t-test} because, being non-parametric, it makes no assumption on the data distribution \cite{woolson2005wilcoxon}. 

In addition, again following established standards \cite{arcuri_practical_2011}, when the $p$-value is $\leq 0.05$, we complement the results obtained by the Wilcoxon test with the Vargha-Delaney $A_{12}$ effect size \cite{vargha2000critique} to assess the size of the obtained difference. This test measures the proportions of items whose difference is higher than zero. Following previous work \cite{sarro_multi-objective_2016}, we consider the effect size \textit{large} if $A_{12} \geq 0.72$, \textit{medium} if $0.64 < A_{12} < 0.72$, and \textit{small} if $A_{12} \le 0.64$.

\section{Results}\label{sec:results}
In this section, we discuss the results of our empirical evaluation. For all RQs (apart from RQ$_{1.1}$, which is a sanity check on the \lifemap domain model), we counted the number of times in which each optimal configuration provided models that are syntactically and semantically better than the baseline. In particular, following previous empirical work \cite{hort_multi-objective_2023}, we use the so-called \textit{Win / Tie / Loss} strategy. We count the number of times a solution scored a Wilcoxon's $p$–value$<0.05$ (Win), $p$–value$>0.99$ (Loss), and $0.05\leq$ $p$–value $\geq0.99$ (Tie). In addition, we count the number of times the obtained $A_{12}$ effect size is \textit{large/medium/small} for each \textit{Win} case. 

\begin{table}[tb!]
    \centering
    \caption{Win / Tie / Loss between optimal solutions and the baseline model}
    \label{tab:win_tie_loss}
    \begin{subtable}{\linewidth}
        \centering
   \begin{tabular}{l|c|c||c|c||c|c}
    \toprule
    & \multicolumn{2}{c||}{\textbf{RQ$_{1.2}$: Zero-Shot}} & \multicolumn{2}{c||}{\textbf{RQ$_{2.1}$: Few-Shot}} & \multicolumn{2}{c}{\textbf{RQ$_{2.2}$: CoT}} \\\midrule
    & \textbf{Cosine} & \textbf{BERT}  & \textbf{Cosine} & \textbf{BERT} & \textbf{Cosine} & \textbf{BERT} \\
    \midrule\midrule
    \textbf{S0} & 4 / 0 / 6 & 2 / 2 / 6 & 5 / 0 / 5 & \cellcolor{gray!25}\underline{5 / 1 / 4} & \cellcolor{gray!25}\underline{7 / 1 / 2} & 4 / 0 / 6 \\
    \textbf{S1} & 1 / 0 / 9 & 2 / 1 / 7 & \cellcolor{gray!25}\underline{8 / 0 / 2} & 4 / 0 / 6 & \cellcolor{gray!25}\underline{6 / 1 / 3} & 1 / 0 / 9 \\
    \textbf{S2} & 1 / 1 / 8 & 1 / 0 / 9 & \cellcolor{gray!25}\underline{6 / 1 / 3} & 4 / 2 / 4 & \cellcolor{gray!25}\underline{6 / 0 / 4} & 1 / 0 / 9 \\
    \textbf{S3} & 3 / 0 / 7 & 1 / 2 / 7 & \cellcolor{gray!25}\underline{9 / 0 / 1} & 2 / 0 / 8 & \cellcolor{gray!25}\underline{8 / 0 / 2} & 5 / 0 / 5 \\
    \textbf{S4} & 1 / 1 / 8 & 0 / 0 / 10 & \cellcolor{gray!25}\underline{6 / 1 / 3} & 3 / 0 / 7 & \cellcolor{gray!25}\underline{6 / 2 / 2} & 3 / 0 / 7 \\
    \textbf{S5} & 4 / 0 / 6 & 3 / 3 / 4 & \cellcolor{gray!25}\underline{7 / 0 / 3} & 3 / 0 / 7 & \cellcolor{gray!25}\underline{7 / 0 / 3} & 2 / 1 / 7 \\\midrule\midrule
    \textbf{$A_{12}$} & 14 / 0 / 0 & 9 / 0 / 0 & 41 / 0 / 0 & 21 / 0 / 0 & 40 / 0 / 0 & 16 / 0 / 0 \\
    \bottomrule
    \end{tabular}
        \caption{Comparison between any optimal solution and the baseline model in all domain models. 
        }
    \label{tab:config_stats}
    \end{subtable}
    \begin{subtable}{\linewidth}
    \centering
\begin{tabular}{l|c|c||c|c||c|c}
        \toprule
        & \multicolumn{2}{c||}{\textbf{RQ$_{1.2}$: Zero-Shot}} & \multicolumn{2}{c||}{\textbf{RQ$_{2.1}$: Few-Shot}} & \multicolumn{2}{c}{\textbf{RQ$_{2.2}$: CoT}} \\ \midrule
         & \textbf{Cosine} & \textbf{BERT}& \textbf{Cosine} & \textbf{BERT}& \textbf{Cosine} & \textbf{BERT} \\
        \midrule
        \midrule
            \textbf{HBMS} & \cellcolor{gray!25}\underline{3 / 1 / 2}& 3 / 0 / 3 & \cellcolor{gray!25}\underline{4 / 0 / 2} & \cellcolor{gray!25}\underline{3 / 1 / 2} & \cellcolor{gray!25}\underline{5 / 0 / 1} & 1 / 1 / 4 \\
            \textbf{OTS} & 0 / 0 / 6 & 0 / 1 / 5 & 3 / 0 / 3 & 0 / 0 / 6 & \cellcolor{gray!25}\underline{4 / 0 / 2} & 2 / 0 / 4 \\
            \textbf{SHAS} & 1 / 0 / 5 & 1 / 0 / 5 & 3 / 0 / 3 & 3 / 0 / 3 & 1 / 2 / 3 & 0 / 0 / 6 \\
            \textbf{BTMS} & 2 / 0 / 4 & 3 / 2 / 1 & \cellcolor{gray!25}\underline{3 / 1 / 2} & 2 / 0 / 4 & \cellcolor{gray!25}\underline{4 / 0 / 2} & \cellcolor{gray!25}\underline{4 / 0 / 2} \\
            \textbf{LabTracker} & 1 / 0 / 5 & 0 / 0 / 6 & \cellcolor{gray!25}\underline{6 / 0 / 0} & \cellcolor{gray!25}\underline{4 / 1 / 1} & \cellcolor{gray!25}\underline{5 / 0 / 1} & 2 / 0 / 4 \\
            \textbf{CelO} & \cellcolor{gray!25}\underline{5 / 0 / 1}& 0 / 2 / 4 & \cellcolor{gray!25}\underline{6 / 0 / 0} & 3 / 0 / 3 & \cellcolor{gray!25}\underline{5 / 0 / 1} & 3 / 0 / 3 \\
            \textbf{TileO} & 0 / 0 / 6 & 0 / 1 / 5 & 2 / 1 / 3 & 0 / 0 / 6 & 3 / 1 / 2 & 0 / 0 / 6 \\
            \textbf{H2S} & 2 / 1 / 3 & 2 / 1 / 3 & \cellcolor{gray!25}\underline{6 / 0 / 0} & 1 / 0 / 5 & \cellcolor{gray!25}\underline{5 / 0 / 1} & 2 / 0 / 4 \\
            \textbf{TSS} & 0 / 0 / 6 & 0 / 1 / 5 & \cellcolor{gray!25}\underline{5 / 0 / 1} & \cellcolor{gray!25}\underline{5 / 0 / 1} & \cellcolor{gray!25}\underline{4 / 1 / 1} & 2 / 0 / 4 \\
            \textbf{Block} & 0 / 0 / 6 & 0 / 0 / 6 & 3 / 0 / 3 & 0 / 1 / 5 & \cellcolor{gray!25}\underline{4 / 0 / 2} & 0 / 0 / 6 \\
        \bottomrule
        \end{tabular}
         \caption{Comparison between all optimal solutions and the baseline model in each domain model.}
    \label{tab:domain_model}
\end{subtable}
\end{table}

Table \ref{tab:config_stats} reports the statistics for each optimal solution, while Table \ref{tab:domain_model} reports the results for each domain model. Note how, in Table \ref{tab:domain_model}, the effect sizes are not reported since they are the same as in Table \ref{tab:config_stats}. In both tables, we highlight cases where the number of wins is higher than losses.

\subsection{RQ$_1$: Zero-Shot Prompting}


\subsubsection{RQ$_{1.1}$: Sanity check}

\begin{figure}[tb]
    \centering
    \includegraphics[width=0.5\linewidth]{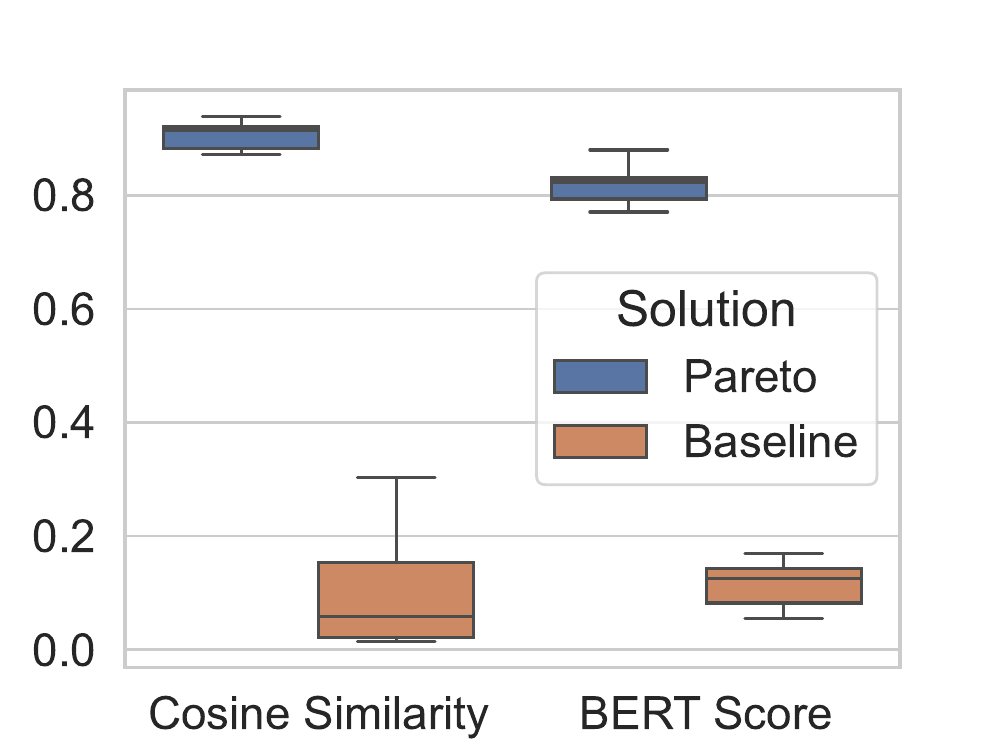}
    \caption{RQ$_{1.1}$: Distribution of Cosine Similarity and BERTScore of models generated by Llama3.1 using the optimal configurations and with standard hyperparameters}
    \label{fig:rq1}
\end{figure}

Figure \ref{fig:rq1} reports the distribution of Cosine Similarity and BERT Score of models generated by Llama3.1 for the \lifemap domain model. The left-most blue boxes show the results obtained by Llama3.1 with the optimal configurations, while the right-most orange boxes show the results obtained by the baseline model with default hyperparameters. As shown in the plot, the syntactical and semantical quality of domain models generated by the LLM with optimal hyperparameters significantly overcomes the results obtained by the baseline. This outcome is confirmed by the results of the Wilcoxon test and the $A_{12}$ effect size, which reported a large significant difference.


The results obtained motivate the need for tuning the LLM's hyperparameter for domain modeling generation. Next, we investigate how the obtained results can be generalized to different domain models.

\subsubsection{RQ$_{1.2}$: Solutions Generalizability}

The results reported in Table \ref{tab:config_stats} for Zero-Shot show how no configuration achieved more wins than losses. This result is confirmed for both Cosine Similarity and BERT Score. However, we also observe how, for the winning cases, the effect size is always \textit{large}. Thus, when the LLM with optimal configurations generates better models, they are significantly better.


Instead, we observe in Table \ref{tab:domain_model} a disagreement in the syntactical and semantical quality of the generated domain models. In fact, under Cosine Similarity (\ie syntactical quality), there are two domain models with more wins than losses - namely \textbf{HBMS} and \textbf{CelO}. These domains pertain to \textit{Hotel Booking Management} and \textit{Celebration Organization}, respectively, and have an average number of classes, attributes, and relations compared to the other domain models. Thus, we found no particular pattern among these use cases and with the \lifemap domain model. Instead, under the BERT Score (\ie semantical quality), we found no domain models with more wins than losses.

In general, this evaluation may lead us to conclude that the optimal solutions returned by our search can not be generalized to different domain models. This result again motivates our next research question, which involves including prompt engineering strategies in the analysis.

\begin{rqanswer}
    \textbf{Answer to RQ$_1$:} The combination of identified optimal solutions and zero-shot prompting can significantly improve the quality of models generated for the training domain. However, the optimal solutions can not be generalized to different domain models.
\end{rqanswer}

\subsection{RQ$_2$: Prompt Engineering}


\subsubsection{RQ$_{2.1}$: Few-Shot}




From the statistics in Table \ref{tab:config_stats}, we observe a discrepancy between the syntactical and semantic similarities of the generated models. Five out of six solutions have a higher count of wins concerning syntactical similarity. In contrast, we identified only one solution that had more wins with respect to the BERT Score. Notably, this solution is also the only one that did not have more wins in terms of Cosine Similarity. This finding may suggest a systematic disagreement between these two metrics when evaluating the quality of domain models. As for the RQ$_1$, we observe how the effect size for the winning cases is always large.

By looking at the statistics for each domain model in Table \ref{tab:domain_model}, we observe how the number of domain models with more wins than losses increases compared with RQ$_{1.2}$. In addition, note how the two winning domain models from RQ$_{1.2}$ - \ie \textbf{HBMS} and \textbf{CelO} - are also winning here.

It is worth noting that the two metrics are concordant in \textbf{LabTracker}, \textbf{HBMS}, and \textbf{TSS} domain models. On the one hand, this is quite expected in the first case since the \textbf{LabTracker} domain is the closest to the \lifemap model, \ie both are tailored for the medical domain even though the latter is more complex. On the other hand, \textbf{HBMS} and \textbf{TSS} are completely different from the medical concepts, even though both contain a lot of generalizations for representing different types of travelers and players, respectively.  

\subsubsection{RQ$_{2.2}$: Chain-of-Thoughts}




By looking at the Win/Tie/Loss scores in Table \ref{tab:config_stats}, we can see how all optimal solutions provide more wins than losses concerning Cosine Similarity. On the contrary, no solution achieves more wins concerning BERT Score. Again, this result may suggest a disagreement of those metrics. 

By instead observing the stats for domain models in Table \ref{tab:domain_model}, the number of domain models in which more optimal solutions are better than the baseline increases in Cosine Similarity while decreasing in BERT Score. 
Remarkably, only the \textbf{BTMS} model achieves the agreement between the two metrics. Our intuition is that the reasoning steps introduced in the CoT prompt lead to this result, thus allowing the usage of hyperparameter tuning in completely different domains. 

From this evaluation, we may conclude that few-shot prompting is preferred to increase both syntactical and semantical similarity. At the same time, chain-of-thought may be suggested to obtain the best syntactical similarity of the generated models.

\begin{rqanswer}
    \textbf{Answer to RQ$_2$}: Adding hyperparameter tuning to prompt engineering strategies significantly improves the quality of generated models. Few-shot prompting is suggested to increase both syntactical and semantical similarity. On the contrary, optimal hyperparameters and chain-of-thought prompting obtained the best syntactical similarity of generated models. 
\end{rqanswer}

\section{Threats to Validity}\label{sec:threats}
This section discusses threats that may hamper the results of our study and the adopted mitigations. 

\medskip
\noindent
\textit{Threats to internal validity} concerns two aspects of our experiment, \ie hyperparameter tuning and the employed prompt engineering techniques. Concerning the first aspect, we followed the state-of-the-art process adopted in operational research. In addition, we perform a sanity check to validate the results obtained from the global and local search. Concerning the second, we acknowledge that prompting can introduce unexpected variability during the generation of the domain models. To mitigate these issues, we compute well-known similarity functions to compare the generated domain models with the real ones. In addition, we used the Win/Tie/Loss test to confirm the outcomes of our study. 


\medskip
\noindent
\textit{External validity} is related to the generalizability of our results to other domains or LLMs. On the one hand, the results show that the hyperparameter tuning is not enough to generalize the results obtained with Llama 3 if different domains are considered. On the other hand, well-founded prompt engineering techniques can contribute to limiting this issue. Another issue is related to the used LLMs, i.e., Llama 3 may not be suitable for modeling tasks. In this respect, we opt for using the most popular open-source model even though we plan to extend our study to additional models with different architecture, \eg DeepSeek, Claude, or GPT models.

\section{Conclusion and Future Work}\label{sec:conclusion}


Motivated by the lack of previous studies in hyperparameter tuning for Large Language Models (LLMs) for supporting the generation of domain models, this paper introduced a novel approach that combines hyperparameter tuning and prompt engineering for supporting domain modeling using a single domain model. To this end, we leveraged the NSGA-II and grid search algorithm to identify a list of optimal hyperparameters for a specific LLM, namely Llama 3, and then employed it to generate ten different domain models. Although the results show that the hyperparameter tuning does not achieve adequate results in all the considered domains, the usage of prompt engineering techniques, especially few-shots, contributes to limiting this effect, thus opening future research in this direction.
In future work, we plan to study in detail each hyperparameter, e.g., temperature or Top-p, to understand their impact during the model generation. In addition, we plan to experiment with additional LLMs (\eg GPT or BERT) and similarity metrics to understand their impact on domain modeling. Finally, we can enlarge the dataset to different domain applications to further study the generalizability of the proposed methodology. 

\begin{credits}
\subsubsection{\ackname} This work has been partially supported by Territori Aperti (a project funded by Fondo Territori, Lavoro e Conoscenza CGIL CISL UIL), by "LIFEMAP-Dalla patologia pediatrica alle malattie cardiovascolari e neoplastiche nell’adulto: mappatura genomica per la medicina e prevenzione personalizzata" Traiettoria 3 “Medicina rigenerativa, predittiva e personalizzata” - Linea di azione 3.1  “Creazione di un programma
di medicina di precisione per la mappatura del genoma umano
su scala nazionale” of the Italian Ministry of Health, by the MOSAICO project (Management, Orchestration and Supervision of AI-agent COmmunities for reliable AI in software engineering) that has received funding from the European Union under the Horizon Research and Innovation Action (Grant Agreement No. 101189664) and Project PRIN 2022 PNRR “FRINGE: context-aware FaiRness engineerING in complex software systEms” grant n. P2022553SL.

\subsubsection{\discintname}
The authors have no competing interests to declare that are relevant to the content of this article. 
\end{credits}

\bibliographystyle{splncs04}
\bibliography{bibliography}

\begin{thebibliography}{10}
\providecommand{\url}[1]{\texttt{#1}}
\providecommand{\urlprefix}{URL }
\providecommand{\doi}[1]{https://doi.org/#1}

\bibitem{ahmad_towards_2023}
Ahmad, A., Waseem, M., Liang, P., Fahmideh, M., Aktar, M.S., Mikkonen, T.: Towards {Human}-{Bot} {Collaborative} {Software} {Architecting} with {ChatGPT}. In: Proceedings of the 27th {International} {Conference} on {Evaluation} and {Assessment} in {Software} {Engineering}. pp. 279--285. {EASE} '23, Association for Computing Machinery, New York, NY, USA (Jun 2023). \doi{10.1145/3593434.3593468}, \url{https://doi.org/10.1145/3593434.3593468}, read\_Status: New Read\_Status\_Date: 2024-07-17T07:38:55.084Z

\bibitem{arcuri_practical_2011}
Arcuri, A., Briand, L.: A practical guide for using statistical tests to assess randomized algorithms in software engineering. In: Proceedings of the 33rd {International} {Conference} on {Software} {Engineering}. pp. 1--10. {ICSE} '11, Association for Computing Machinery, New York, NY, USA (May 2011). \doi{10.1145/1985793.1985795}, \url{https://dl.acm.org/doi/10.1145/1985793.1985795}

\bibitem{10651492}
Ardimento, P., Bernardi, M.L., Cimitile, M.: Teaching uml using a rag-based llm. In: 2024 International Joint Conference on Neural Networks (IJCNN). pp.~1--8 (2024). \doi{10.1109/IJCNN60899.2024.10651492}

\bibitem{arora2024optimizinglargelanguagemodel}
Arora, C., Sayeed, A.I., Licorish, S., Wang, F., Treude, C.: Optimizing large language model hyperparameters for code generation (2024), \url{https://arxiv.org/abs/2408.10577}

\bibitem{arulmohan_extracting_2023}
Arulmohan, S., Meurs, M.J., Mosser, S.: Extracting {Domain} {Models} from {Textual} {Requirements} in the {Era} of {Large} {Language} {Models}. In: 2023 {ACM}/{IEEE} {International} {Conference} on {Model} {Driven} {Engineering} {Languages} and {Systems} {Companion} ({MODELS}-{C}). pp. 580--587. IEEE, V\"{a}ster\aa{}s, Sweden (Oct 2023). \doi{10.1109/MODELS-C59198.2023.00096}, \url{https://ieeexplore.ieee.org/document/10350787/}

\bibitem{10.1145/2976767.2976785}
Batot, E., Sahraoui, H.: A generic framework for model-set selection for the unification of testing and learning mde tasks. In: Proceedings of the ACM/IEEE 19th International Conference on Model Driven Engineering Languages and Systems. p. 374–384. MODELS '16, Association for Computing Machinery, New York, NY, USA (2016), \url{https://doi.org/10.1145/2976767.2976785}

\bibitem{10.1145/3640310.3674085}
Ben~Chaaben, M., Ben~Sghaier, O., Dhaouadi, M., Elrasheed, N., Darif, I., Jaoua, I., Oakes, B., Syriani, E., Hamdaqa, M.: Toward intelligent generation of tailored graphical concrete syntax. In: Proceedings of the ACM/IEEE 27th International Conference on Model Driven Engineering Languages and Systems. p. 160–171. MODELS '24, Association for Computing Machinery, New York, NY, USA (2024). \doi{10.1145/3640310.3674085}, \url{https://doi.org/10.1145/3640310.3674085}

\bibitem{10.1145/3567512.3567534}
Bertram, V., Bo\ss{}, M., Kusmenko, E., Nachmann, I.H., Rumpe, B., Trotta, D., Wachtmeister, L.: Neural language models and few shot learning for systematic requirements processing in mdse. In: Proceedings of the 15th ACM SIGPLAN International Conference on Software Language Engineering. p. 260–265. SLE 2022, Association for Computing Machinery, New York, NY, USA (2022). \doi{10.1145/3567512.3567534}, \url{https://doi .org/10.1145/3567512.3567534}

\bibitem{repl_package}
Bulhakov, V., d'Aloisio, G., Di~Sipio, C., Di~Marco, A., Di~Ruscio, D.: {LLama Ecore Study} (Jan 2025), \url{https://github.com/VPLEV23/ER_LLM}

\bibitem{10.1145/3712008}
Burgue\~{n}o, L., Di~Ruscio, D., Sahraoui, H., Wimmer, M.: Automation in model-driven engineering: A look back, and ahead. ACM Trans. Softw. Eng. Methodol.  (Jan 2025), \url{https://doi.org/10.1145/3712008}, just Accepted

\bibitem{camara_assessment_2023}
C\'{a}mara, J., Troya, J., Burgue\~{n}o, L., Vallecillo, A.: On the assessment of generative {AI} in modeling tasks: an experience report with {ChatGPT} and {UML}. Software and Systems Modeling  \textbf{22}(3),  781--793 (Jun 2023). \doi{10.1007/s10270-023-01105-5}, \url{https://doi.org/10.1007/s10270-023-01105-5}

\bibitem{chaaben2023towards}
Chaaben, M.B., Burgue\~{n}o, L., Sahraoui, H.: Towards {Using} {Few}-{Shot} {Prompt} {Learning} for {Automating} {Model} {Completion}. In: Proceedings of the 45th {International} {Conference} on {Software} {Engineering}: {New} {Ideas} and {Emerging} {Results}. pp. 7--12. {ICSE}-{NIER} '23, IEEE, IEEE Press, Melbourne, Australia (Sep 2023). \doi{10.1109/ICSE-NIER58687.2023.00008}, \url{https://dl.acm.org/doi/10.1109/ICSE-NIER58687.2023.00008}

\bibitem{chen_automated_2023}
Chen, K., Yang, Y., Chen, B., Hern\'{a}ndez~L\'{o}pez, J.A., Mussbacher, G., et~al.: Automated {Domain} {Modeling} with {Large} {Language} {Models}: {A} {Comparative} {Study}. In: 2023 {ACM}/{IEEE} 26th {International} {Conference} on {Model} {Driven} {Engineering} {Languages} and {Systems} ({MODELS}). pp. 162--172 (Oct 2023). \doi{10.1109/MODELS58315.2023.00037}, \url{https://ieeexplore.ieee.org/abstract/document/10344012}

\bibitem{pmlr-v139-chen21v}
Chen, P.H., Wei, W., Hsieh, C.J., Dai, B.: Overcoming catastrophic forgetting by bayesian generative regularization. In: Meila, M., Zhang, T. (eds.) Proceedings of the 38th International Conference on Machine Learning. Proceedings of Machine Learning Research, vol.~139, pp. 1760--1770. PMLR (18--24 Jul 2021), \url{https://proceedings.mlr.press/v139/chen21v.html}

\bibitem{10.1145/3640310.3674089}
Costa, C.D., L\'{o}pez, J.A.H., Cuadrado, J.S.: Modelmate: A recommender for textual modeling languages based on pre-trained language models. In: Proceedings of the ACM/IEEE 27th International Conference on Model Driven Engineering Languages and Systems. p. 183–194. MODELS '24, Association for Computing Machinery, New York, NY, USA (2024). \doi{10.1145/3640310.3674089}, \url{https://doi .org/10.1145/3640310.3674089}

\bibitem{d2024exploring}
d'Aloisio, G., Fortz, S., Hanna, C., Fortunato, D., Bensoussan, A., Mendiluze~Usandizaga, E., Sarro, F.: Exploring llm-driven explanations for quantum algorithms. In: Proceedings of the 18th ACM/IEEE International Symposium on Empirical Software Engineering and Measurement. pp. 475--481 (2024)

\bibitem{nsgaII}
Deb, K., Agrawal, S., Pratap, A., Meyarivan, T.: A fast and elitist multiobjective genetic algorithm: {NSGA-II}. {IEEE} Trans. Evol. Comput.  \textbf{6}(2),  182--197 (2002)

\bibitem{deroy2024code}
Deroy, A., Maity, S.: Code generation and algorithmic problem solving using llama 3.1 405b. arXiv preprint arXiv:2409.19027  (2024)

\bibitem{di2022memorec}
Di~Rocco, J., Di~Ruscio, D., Di~Sipio, C., Nguyen, P.T., Pierantonio, A.: Memorec: a recommender system for assisting modelers in specifying metamodels. Software and Systems Modeling pp. 1--21 (2022)

\bibitem{10.1145/3652620.3687808}
Di~Sipio, C., Rubei, R., Di~Rocco, J., Di~Ruscio, D., Iovino, L.: On the use of llms to support the development of domain-specific modeling languages. In: Proceedings of the ACM/IEEE 27th International Conference on Model Driven Engineering Languages and Systems. p. 596–601. MODELS Companion '24, Association for Computing Machinery, New York, NY, USA (2024). \doi{10.1145/3652620.3687808}, \url{https://doi.org/10.1145/3652620.3687808}

\bibitem{10.1145/3709358}
Fan, L., Liu, J., Liu, Z., Lo, D., Xia, X., Li, S.: Exploring the capabilities of llms for code change related tasks. ACM Trans. Softw. Eng. Methodol.  (Dec 2024). \doi{10.1145/3709358}, \url{https://doi.org/10.1145/3709358}, just Accepted

\bibitem{forrest1996genetic}
Forrest, S.: Genetic algorithms. ACM computing surveys (CSUR)  \textbf{28}(1),  77--80 (1996)

\bibitem{10.1162/EVCO_a_00128}
Giagkiozis, I., Fleming, P.J.: Pareto front estimation for decision making. Evolutionary Computation  \textbf{22}(4),  651--678 (2014). \doi{10.1162/EVCO_a_00128}

\bibitem{gong2024greenstableyolo}
Gong, J., Li, S., d’Aloisio, G., Ding, Z., Ye, Y., Langdon, W.B., Sarro, F.: Greenstableyolo: Optimizing inference time and image quality of text-to-image generation. In: International Symposium on Search Based Software Engineering. pp. 70--76. Springer Nature Switzerland Cham (2024)

\bibitem{hamdani2007multi}
Hamdani, T.M., Won, J.M., Alimi, A.M., Karray, F.: Multi-objective feature selection with nsga ii. In: Adaptive and Natural Computing Algorithms: 8th International Conference, ICANNGA 2007, Warsaw, Poland, April 11-14, 2007, Proceedings, Part I 8. pp. 240--247. Springer (2007)

\bibitem{hort_multi-objective_2023}
Hort, M., Moussa, R., Sarro, F.: Multi-objective search for gender-fair and semantically correct word embeddings. Applied Soft Computing  \textbf{133},  109916 (Jan 2023). \doi{10.1016/j.asoc.2022.109916}, \url{https://www.sciencedirect.com/science/article/pii/S1568494622009656}

\bibitem{kora2017crossover}
Kora, P., Yadlapalli, P.: Crossover operators in genetic algorithms: A review. International Journal of Computer Applications  \textbf{162}(10) (2017)

\bibitem{lewis2020retrieval}
Lewis, P., Perez, E., Piktus, A., Petroni, F., Karpukhin, V., Goyal, N., K\"{u}ttler, H., Lewis, M., Yih, W.t., Rockt\"{a}schel, T., Riedel, S., Kiela, D.: Retrieval-augmented generation for knowledge-intensive nlp tasks. In: Proceedings of the 34th International Conference on Neural Information Processing Systems. NIPS '20, Curran Associates Inc., Red Hook, NY, USA (2020)

\bibitem{modelset}
L{\'o}pez, J.A.H., C{\'a}novas~Izquierdo, J.L., Cuadrado, J.S.: Modelset: a dataset for machine learning in model-driven engineering. Software and Systems Modeling pp. 1--20 (2021)

\bibitem{ma2024fine}
Ma, X., Wang, L., Yang, N., Wei, F., Lin, J.: Fine-tuning llama for multi-stage text retrieval. In: Proceedings of the 47th International ACM SIGIR Conference on Research and Development in Information Retrieval. pp. 2421--2425 (2024)

\bibitem{LLama3}
{Meta}: {Meta LLama3}. \url{https://llama.meta.com/llama3/} (2024), \url{https://llama.meta.com/llama3/}

\bibitem{mussbacher_opportunities_2020}
Mussbacher, G., Combemale, B., Kienzle, J., Abrah\~{a}o, S., Ali, H., Bencomo, N., B\'{u}r, M., Burgue\~{n}o, L., Engels, G., Jeanjean, P., J\'{e}z\'{e}quel, J.M., K\"{u}hn, T., Mosser, S., Sahraoui, H., Syriani, E., Varr\'{o}, D., Weyssow, M.: Opportunities in intelligent modeling assistance. Software and Systems Modeling  \textbf{19}(5),  1045--1053 (Sep 2020). \doi{10.1007/s10270-020-00814-5}, \url{http://link.springer.com/10.1007/s10270-020-00814-5}

\bibitem{10.5555/993859}
Rumbaugh, J., Jacobson, I., Booch, G.: Unified Modeling Language Reference Manual, The (2nd Edition). Pearson Higher Education (2004)

\bibitem{10.1145/3417990.3421385}
Saini, R., Mussbacher, G., Guo, J.L.C., Kienzle, J.: Domobot: A bot for automated and interactive domain modelling. In: Proceedings of the 23rd ACM/IEEE International Conference on Model Driven Engineering Languages and Systems: Companion Proceedings. MODELS '20, Association for Computing Machinery, New York, NY, USA (2020), \url{https://doi.org/10.1145/3417990.3421385}

\bibitem{salton1988term}
Salton, G., Buckley, C.: Term-weighting approaches in automatic text retrieval. Information processing \& management  \textbf{24}(5),  513--523 (1988)

\bibitem{sarro_multi-objective_2016}
Sarro, F., Petrozziello, A., Harman, M.: Multi-objective software effort estimation. In: Proceedings of the 38th {International} {Conference} on {Software} {Engineering}. pp. 619--630. {ICSE} '16, Association for Computing Machinery, New York, NY, USA (May 2016). \doi{10.1145/2884781.2884830}, \url{https://dl.acm.org/doi/10.1145/2884781.2884830}

\bibitem{vargha2000critique}
Vargha, A., Delaney, H.D.: A critique and improvement of the cl common language effect size statistics of mcgraw and wong. Journal of Educational and Behavioral Statistics  \textbf{25}(2),  101--132 (2000)

\bibitem{wei2022chain}
Wei, J., Wang, X., Schuurmans, D., Bosma, M., Xia, F., Chi, E., Le, Q.V., Zhou, D., et~al.: Chain-of-thought prompting elicits reasoning in large language models. Advances in neural information processing systems  \textbf{35},  24824--24837 (2022)

\bibitem{weyssow2022recommending}
Weyssow, M., Sahraoui, H., Syriani, E.: Recommending metamodel concepts during modeling activities with pre-trained language models. Software and Systems Modeling pp. 1--19 (2022)

\bibitem{wohlin_experimentation_2012}
Wohlin, C., Runeson, P., Höst, M., Ohlsson, M.C., Regnell, B., Wesslén, A.: Experimentation in {Software} {Engineering}. Springer Berlin Heidelberg, Berlin, Heidelberg (2012). \doi{10.1007/978-3-642-29044-2}, \url{http://link.springer.com/10.1007/978-3-642-29044-2}

\bibitem{woolson2005wilcoxon}
Woolson, R.F.: Wilcoxon signed-rank test. Encyclopedia of Biostatistics  \textbf{8} (2005)

\bibitem{zhang2019bertscore}
Zhang, T., Kishore, V., Wu, F., et~al.: Bertscore: Evaluating text generation with bert. arXiv preprint arXiv:1904.09675  (2019)

\end{thebibliography}

\end{document}